# 煤炭企业经营状况与股票收益率的非同步波动

乔柯南

（中国科学院管理决策与信息系统重点实验室，北京 100190）

**摘要**：为研究煤炭企业经营状况和金融市场股票收益率波动之间的相关性，本文采用时差相关分析和 PageRank 方法，对 A 股市场 21 只煤炭行业相关股票进行了收益率的领先滞后关系研究，进而对个股的领先强度进行了打分评价。基于 2011 年金融市场的相关数据，本文揭示了煤炭行业相关股票收益率波动的非同步性。对收益率波动的传导路径进行了分层。将所得结果和 2011 年各企业经营状况进行了对比研究，讨论了两者之间的联系。结果表明，规模较大且经营状况良好的煤炭企业，其股票收益率往往有更好的先行性；PageRank 方法得到的打分结果和公司规模以及经营状况均存在显著的正相关关系。

关键词：煤炭行业，股票，收益率，时差相关分析，PageRank

中图分类号：F830.9      文献标识码：      文章编号：

## Coal Enterprise Management and Asynchronism of Return

Qiao Kenan

（Key Laboratory of Management Decision and Information Systems, Chinese Academy of Sciences, Beijing 100190）

**Abstract**: For researching the association between coal enterprise management and return in financial market, this paper applies the method of time difference relevance and PageRank to seek the leader-index of a stock set containing 21 coal enterprises in A-share market and score those stocks. Based on the return in 2011, the asynchronism of the return series is revealed and presents a hierarchical structure of our stock set. Finally, we compare the result with the firm-level variables and discuss the relation between them. The results show that those large coal enterprises with a good management condition always present an antecedence of stock return; there is a significant positive association between company scale and the score given by PageRank method.

  企业经营状况和金融市场收益率的波动性一直是学术界关注的热点问题之一。本文通过改进的 PageRank 方法对它们进行了分析和讨论。2011 年是国际经济形势持续动荡的一年，不明朗的经济环境势必影响企业的经营状况。煤炭行业作为我国重要的能源支柱不可避免地受到了一定冲击。宏观信息和企业经营状况是金融市场上投资者进行交易的主要参考依据之一。本文运用改进的 PageRank 方法对 21 只'煤炭股'进行了分层，探讨了收益率波动的领先滞后关系，并对比企业经营状况研究了个股对宏观信息的反馈能力。

# 1 研究方法介绍

## 1.1 时差相关分析

时差相关分析是宏观经济景气分析中经常采取的计量方法，是用以判断相关指标的先行滞后关系进而对指标进行筛选的一种技术[1]。以下简述这一方法：

$x = \{x_n\}_{n=1}^N, y = \{y_n\}_{n=1}^N$ 为两个时间序列，假设 $x_n$ 领先于 $y_n$，对任意自然数 $h$，

$$r^{x,y}(h) = \frac{\sum_{n=h+1}^{N}(x_{n-h}-\bar{x}_{-h})(y_n-\bar{y}_h)}{\sqrt{\sum_{n=h+1}^{N}(x_{n-h}-\bar{x}_{-h})^2 \sum_{n=h+1}^{N}(y_n-\bar{y}_h)^2}},$$

称为 $x$ 关于 $y$ 领先 $h$ 期的时差相关系数，其中 $\bar{x}_{-h} = \frac{\sum_{n=h+1}^{N} x_{n-h}}{N-h}$，$\bar{y}_h = \frac{\sum_{n=h+1}^{N} y_n}{N-h}$。进一步，取 $h^* > 0$ 使得 $r^{x,y}(h^*) = \max_{h>0}|r^{x,y}(h)|$，$h^*$ 被认为是 $x_n$ 领先于 $y_n$ 的期数，而 $r^{x,y}(h^*) = r^{x,y}$ 则被认为是 $x_n$ 领先于 $y_n$ 的时差相关系数。如果 $|r^{x,y}| > |r^{y,x}|$ 则认为 $x_n$ 领先于 $y_n$.

## 1.2 PageRank 方法

时差相关分析可以有效地分析两个时间序列间的领先滞后关系，但更多时候我们需要考虑多个时间序列间的领先滞后关系。这时候就必须推广上述方法，使其适应于多个时间序列的领先滞后关系研究。下面介绍一种可能的方法，PageRank 方法。

PageRank 方法最初被 google 用来对互联网上的海量网页进行打分评价[2-5]，后来被一些学者用于从多个时间序列组成的指标集中萃取领先指标（Leader Index）[6]。下面简述这一方法以及对此方法的一些改进。

考虑多个时间序列组成的指标集 $S = \{\{s_t^n\}_{t=1}^T | n=1,2,...,N\}$ 上标 $n$ 代表指标的编号，下标 $t$ 为观测时间点。对任意指标集中的一对指标 $s_t^n$ 和 $s_t^m$ 进行时差相关分析，得到 $s_t^n$ 领先于 $s_t^m$ 的领先期为 $h$ 的时差相关系数 $r^{s^n,s^m}(h)$. 在某种权重下对各个领先期下的相关系数进行加权求和，得到 $s_t^n$ 关于 $s_t^m$ 的领先强度

$$r^{s^n,s^m} = \sum_{h=1}^{H} r^{s^n,s^m}(h)/H \cdot$$

注意这里的定义和 1.1 中有所不同，这样的处理主要是考虑到很多情况下指标的领先期不是固定不变的。而平均化的处理会降低由于领先期的浮动带来的不稳定性。

接下来根据实际经验设置阈值 $l > 0$. 如果 $|r^{s^n,s^m}| > \lambda$，定义 $W_{n,m} = |r^{s^n,s^m}|$，否则 $W_{n,m} = 0$. 这样，以各个指标 $s_t^n$ 为顶点，$\mathbf{W} = (W_{n,m})_{n,m}$ 为边的权重矩阵就得到了一个赋权有向图 $G = (S, \mathbf{W})$.

PageRank 方法的核心在于之后的循环打分环节。考虑下面的迭代序列，

$$score^{(k+1)} = (1-a)e_{N,1} + a\mathbf{H} \cdot score^{(k)}, k \in \mathbb{N},$$

这里 $\alpha > 0$ 是一个给定的阻尼系数，$e_{N,1}$ 为 $N$ 阶各分量全为 1 的列向量，$\boldsymbol{H}$ 是 $\boldsymbol{W}$ 按列进行归一化得到的矩阵。$\boldsymbol{score}^{(k)}$ 各个分量为对应节点指标的打分，$\boldsymbol{score}^{(0)}$ 为事先给定的初始打分向量。可以证明当 $\alpha$ 充分小时，迭代序列必将收敛到一个和初始打分 $\boldsymbol{score}^{(0)}$ 无关的向量 $\boldsymbol{score}^{(+\infty)}$（参文献[7]）。事实上，

$$\boldsymbol{score}^{(+\infty)} = (1-a)(\boldsymbol{I}-a\boldsymbol{H})^{-1}\boldsymbol{e}_{N,1},$$

只需让 $\alpha \boldsymbol{H}$ 的谱半径小于 1 即可。而 $\boldsymbol{score}^{(+\infty)}$ 各分量就是最终得到的各节点的打分。

这里指出，由于很多经济问题所考虑的时间序列和网页打分有一定区别，序列的双向因果关系是及其常见的。所以这里并没有完全照搬文献[6]的方法，而是允许出现两指标间的双向因果关系，即存在 $W_{n,m}$ 和 $W_{m,n}$ 均大于 0 的可能。另外由于对权重矩阵做了这样的宽松化处理，我们以下在文献[6]的基础上对领先指标的萃取方法做相应的修正。

首先对所有指标按打分从高到低排列 $s_t^{n_1}, s_t^{n_2},...,s_t^{n_N}$，对各个指标依次进行检查。对指标 $s_t^i$ 而言，如果

$$\sum_{u \in U} score_u^{+\infty} \cdot \frac{H_{u,i}}{\sum_k H_{u,k}} \geq score_i^{+\infty},$$

这里 $U$ 为一切尚未被删除的指标集。则删除指标 $s_t^i$. 当所有指标均被检查之后，余下未被删除的指标就是整个指标集的领先指标。

由上述方法得到的领先指标可以看作整个指标集最先接收到'信息'的子集，将它标记为第一层，并从指标集中将其删去。将余下指标视为新的指标集，重新进行时差相关分析和 PageRank 打分，可以得到另一组领先指标，作为整个指标集的第二层信息接收子集。依次类推便可以对指标集进行分层结构化。

## 2 实证研究

### 2.1 数据选取

2011 年我国煤炭企业 100 强中共拥有 41 家上市公司。其中冀中能源集团和中煤能源集团分别拥有 3 家上市公司；中平能化集团、山西焦煤集团和阳泉煤业集团分别拥有 2 家上市公司；新增 3 家上市企业分别是淮南矿业集团、义马煤业集团和山东矿机集团。所属行业为煤炭选采业的深沪两市 A 股上市公司共计 26 家。同一集团所属的上市公司只选取一只个股作为代表。剔除因停牌时间过长的郑州煤电和靖远煤电，余下共 21 支股票如下：中国神华，冀中能源，平煤股份，西山煤电,中煤能源，开滦股份，潞安环能，阳泉煤业，大同煤业，兖州煤业，山煤国际，露天煤业，神火股份，恒源煤电，昊华能源，安源股份，兰花科创，盘江股份，国投新标，平庄能源，大有能源。

我们将运用时差相关分析和 PageRank 方法，依据全年所有有效交易日（非节假日且股

票未停牌的交易日）的对数收益率日度数据对它们进行分层结构化。记股票 $n$ 在第 $t$ 个有效交易日的对数收益率为

$$r_t^n = \log(VWAP_t^n) - \log(VWAP_{t-1}^n),$$

这里 $VWAP_t^n$ 为股票 $n$ 在第 $t$ 个有效交易日的均价，由交易所每日收盘时公布的数据得到。

另外，我们选取上述公司的总资产，年度营业收入，净利润和利润总额作为企业经营状况的度量指标。以便对比研究。

表 1  企业经营状况指标

Table 1 Firm-level variables

| 证券简称 | 资产总计/万元 | 营业收入/万元 | 净利润/万元 | 利润总额/万元 |
|---|---|---|---|---|
| 中国神华 | 33926800 | 1520630 | 3718700 | 5330400 |
| 冀中能源 | 2522857 | 3028916 | 239616 | 332108 |
| 平煤股份 | 1842066 | 2290456 | 184991 | 236509 |
| 西山煤电 | 2895275 | 169435 | 264437 | 383696 |
| 中煤能源 | 12081508 | 7126842 | 690898 | 1022161 |
| 开滦股份 | 1605333 | 1515407 | 86885 | 114392 |
| 潞安环能 | 2936369 | 2142768 | 343673 | 412881 |
| 泉阳煤业 | 2225847 | 2794063 | 241246 | 332102 |
| 大同煤业 | 1696706 | 1045381 | 128700 | 281800 |
| 兖州煤业 | 7282854 | 2484439 | 900862 | 1211383 |
| 山煤国际 | 2165550 | 3864418 | 75577 | 196320 |
| 露天煤业 | 746356 | 567233 | 145559 | 180056 |
| 神火股份 | 2466769 | 1690263 | 115884 | 152147 |
| 恒源煤电 | 1181217 | 699025 | 95107 | 12733 |
| 昊华能源 | 883794 | 404717 | 87256 | 117460 |
| 安源煤业 | 222757 | 119907 | 4244 | 9940 |
| 兰花科创 | 1356165 | 581157 | 131423 | 171423 |
| 盘江股份 | 829375 | 546924 | 134522 | 159216 |
| 国投新集 | 1945922 | 700853 | 125109 | 171421 |
| 平庄能源 | 519028 | 322624 | 64775 | 76074 |
| 大有能源 | 36022 | 24296 | 509 | 2109 |

注： 以上取自中国煤炭工业协会《2011 年中国煤炭企业 100 强分析报告》中公布的有关数据。

## 2.2 实证结果和对比分析

将样本 $\{r_t^n\}_{t,n}$ 看做 $n$ 个时间序列组成的指标集，对其进行时差相关分析和 PageRank 打分。将阻尼系数取为 0.85（这也是 google 采取的数值），为保证结果的可靠性，作者另外尝试将阻尼系数取为 0.83, 0.84, 0.86, 0.87，而最终得到的指标分层结果并未出现明显改变。此外，我们设定相关系数阈值为 0.1。这里设定的阈值是非常小的，一方面因为这里采

取了多先行期平均的方式提取时差相关系数，这样时差相关系数普遍会比较小。同时较小的阈值也最大限度地保留了时序相关的信息。另外取

$$r^{s^n,s^m} = \sum_{t>0} r^{s^n,s^m}(h) \cdot \frac{d_h}{\text{sum}(d)},$$

这里

$d_h = \frac{1}{1+|h-t_0|}$, $t = \arg\max_{h>0} r^{s^n,s^m}(t)$. 这样的处理一是为了最大限度地获取先行指标在最优领先期携带的信息，同时也平衡了先行期的非稳定性，实践证明效果较好。利用 PageRank 方法和 1.2 中分层结构化技术可以得到的各股的最终打分以及分层结构如下：

表 2  PageRank 打分结果

Table 2　PageRank score

| 证券名称 | 中国神华 | 冀中能源 | 平煤股份 | 西山煤电 | 中煤能源 |
|---|---|---|---|---|---|
| 打分 | 1.6637 | 1.3142 | 0.754 | 1.2397 | 2.0636 |
| 证券名称 | 开滦股份 | 潞安环能 | 阳泉煤业 | 大同煤业 | 兖州煤业 |
| 打分 | 1.2397 | 1.0342 | 0.6651 | 0.8353 | 1.6961 |
| 证券名称 | 山煤国际 | 露天煤业 | 神火股份 | 恒源煤电 | 昊华能源 |
| 打分 | 0.9742 | 0.9321 | 1.6541 | 0.4401 | 1.0396 |
| 证券名称 | 安源股份 | 兰花科创 | 盘江股份 | 国投新集 | 平庄能源 |
| 打分 | 0.4178 | 0.4062 | 0.356 | 0.8299 | 1.4444 |
| 证券名称 | 大有能源 | | | | |
| 打分 | 0.3252 | | | | |

表 3 股票集的分层结构

Table 3　Hierarchical structure of the stock set

| 层次 | 证券名称 |
|---|---|
| 1 | 中国神华，中煤能源，兖州煤业，冀中能源 |
| 2 | 平煤股份，神火股份，国投新集，西山煤电，大同煤业，开滦股份，山煤国际，潞安环能，露天煤业，昊华能源， |
| 3 | 阳泉煤业，兰花科创,恒源煤电，平庄能源， |
| 4 | 大有能源，盘江股份，安源股份 |

从上表我们看到，所在层次相对高的公司规模均相对较大，实力较强，社会影响力大。为做更深入的量化比较研究，以下将每一层中所有公司的资产总额，年度营业收入，净利润和利润总额分别在层内做平均，可以得到下面的对比结果：

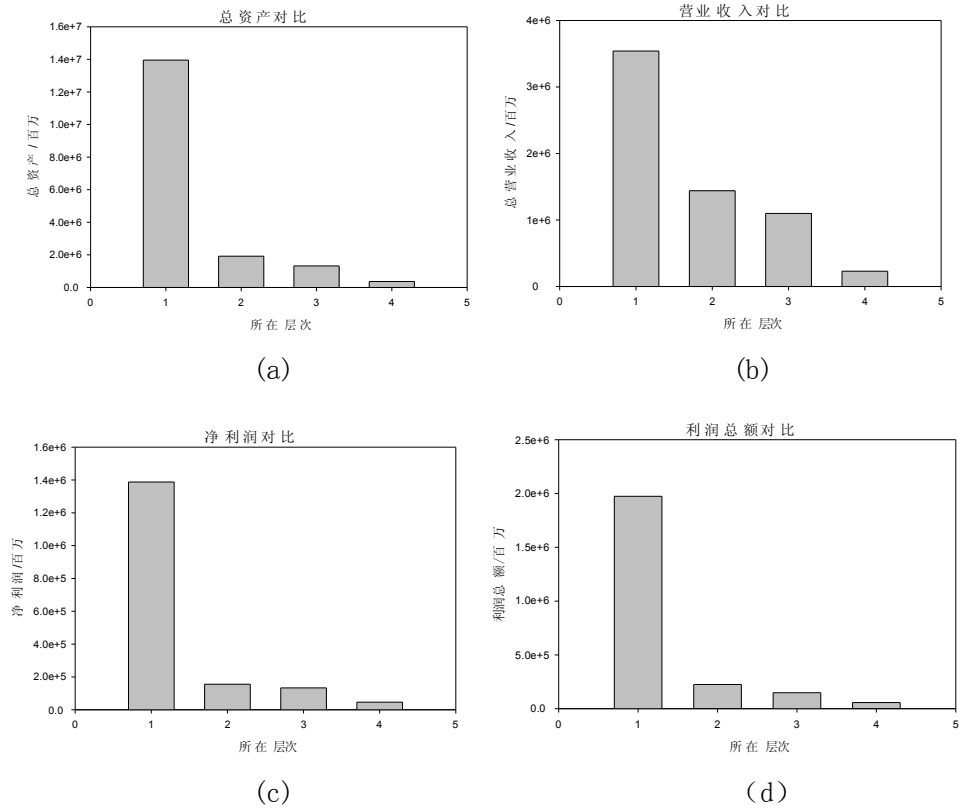

**图 1 企业经营状况与分层结构**

Fig.1 Firm-level variables and hierarchical structure

一般而言，规模较大且经营状况良好的煤炭企业有更好的信息披露能力[9-11]。这些煤炭企业是整个行业的'领军企业'，他们的决策会影响到整个行业的未来发展。从上述柱状图可见，这些信息也是金融市场上投资者参考的重要因素。实力较强的煤炭企业做出的重要决策会被更快地披露到金融市场上，促使投资者进行交易或改变交易策略。而其它相关股票的价格调整是相对滞后的。最后我们对 PageRank 打分和企业经营状况间的关系进行一个简要的分析。将各只股票的 PageRank 打分和该股票发行公司的总资产，年度营业收入，净利润和利润总额的对数值分别做散点图。结果如下：

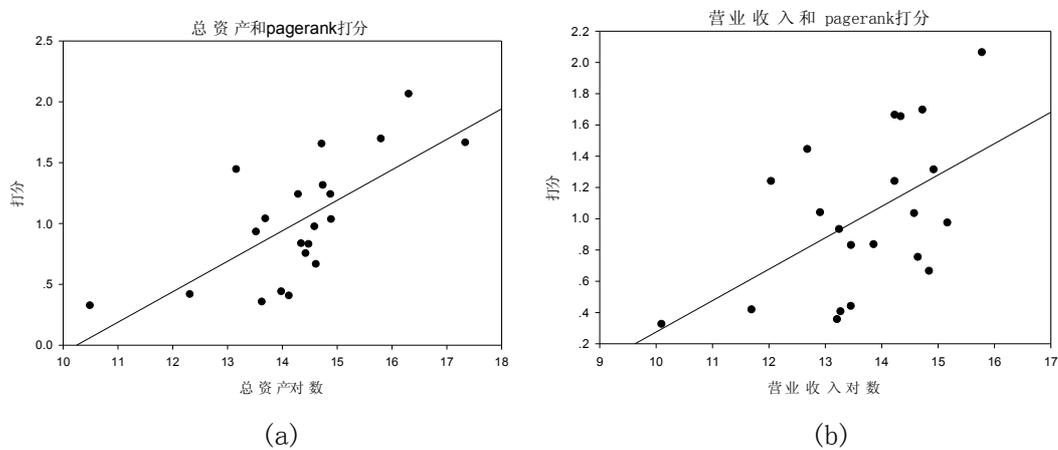

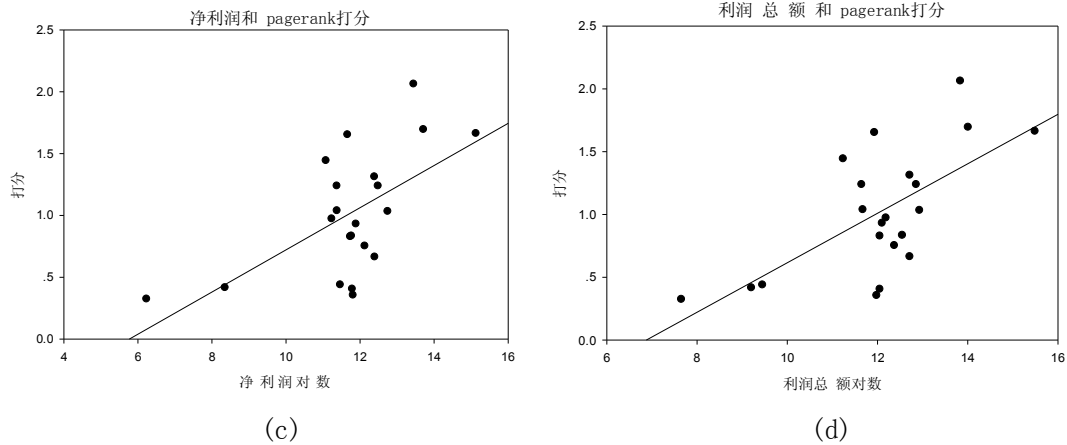

(c)　　　　　　　　　　　　　　(d)

图 2　企业经营状况和 PageRank 打分

Fig.2 Firm-level variables and PageRank score

黑线是最小二乘拟合得到的趋势线。为检验拟合线趋势是否显著，我们估计下面的回归方程

$$score^{(+\infty)} = \alpha + \beta \cdot x + \varepsilon : x = 资产总计对数,营业收入对数,净利润对数,利润总额对数。$$

回归系数 $\beta$ 的显著性检验结果如下，

**表3　$\beta$ 的显著性检验**

**Table 3　Significance test of $\beta$**

| $x$ 取值 | 系数估计 | 标准误 | t统计量 | p值 | 回归方程 $R^2$ |
| --- | --- | --- | --- | --- | --- |
| 资产总计对数 | 0.25 | 0.06 | 4.17 | 0.00 | 0.48 |
| 营业收入对数 | 0.20 | 0.07 | 2.74 | 0.01 | 0.28 |
| 净利润对数 | 0.17 | 0.05 | 3.34 | 0.00 | 0.37 |
| 利润总额对数 | 0.20 | 0.05 | 3.86 | 0.00 | 0.44 |

由此结果可见，PageRank 打分结果与总资产，年度营业收入，净利润和利润总额均存在显著的正相关关系。而 PageRank 打分所要评价的目标是某一指标在整个指标集中的领先强度，打分越高表示该指标在指标集中领先性越强。这样的结论和前面的分析结果也是相吻合的。

## 3 结 论

本文运用改进的时差相关分析和 PageRank 方法，对两市 A 股市场 21 只煤炭行业相关股票对数收益率进行领先滞后关系研究，实现了领先指标的萃取和分层结构化。同时对比企业经营状况对结果进行了讨论和解释。结果表明：实力较强且经营状况良好的大型煤炭企业所发行的股票，其收益率的波动往往较其他煤炭企业有更好的领先性，在分层结构中处于比较

靠前的位置。可见实力较强的大型煤炭企业在金融市场上能更为及时地反映价格波动的信息，对宏观消息的传导和披露能力也更强。另一方面，PageRank 方法给出的打分关于企业资产总额，年度营业收入，净利润和利润总额均呈现正相关的关系。这一结论也说明在金融市场上个股收益率波动的领先性和企业的实力成正比。

**参考文献：**

[1] 董文泉,高铁梅.经济周期波动的分析与预测方法[M].长春:吉林大学出版社，1998：434-482.
[2] LANGVILLE A, MEYER C. Google's PageRank and Beyond: The Science of Search Engine Ranking[M]. Princeton University Press, 2006.
[3] GRECO G, ZUMPANO E. A probabilistic approach for distillation and ranking of web pages[J]. World Wide Web, 2001, 4(3): 89–207.
[4] BRIN S, PAGE L. The anatomy of a large-scale hypertextual web search engine[J]. Computer Network ISDN System, 1998, 30(1–7): 107–117.
[5] PAGE L, BRIN S, MOTWANI R, WINOGRAD T. The PageRank citation ranking: Bringing order to the web[EB/OL]. 1999[2001-10-30]. http://ilpubs.stanford.edu:8090/422/1/1999-66.pdf.
[6] WU Di, KE Yi-ping, Jeffrey, Philip S, LEI Chen. Leadership discovery when data correlatively evolve[J]. World Wide Web,2011,14(1): 1-25.
[7] ALFIC QUARTERONI, RICCARDO SACCO, FAUSTO SALERI. Numerical Mathematics[M]. Springer-Verlag, 2000: 126-129.
[8] GROSSMAN, STIGLITZ. Information and Competition Price System[J]. The American Economic Review, 1976, 66(2): 246-253.
[9] HOOGHIEMSTRA R. Corporate Communication and Impression Management-New Perspectives Why Companies Engage in Corporate Social Reporting[J]. Journal of Business Ethics, 2000, 27(1-2): 55–68.
[10] BOWMAN, EDWARD H. Strategy, Annual Reports and Alchemy[J]. California Management Review, 1978, 20(3): 64-71.
[11] LEV B, Penman S. Voluntary Forecast Disclosure, Non—disclosure, and Stock Prices[J]. Journal of Accounting Research，28(1): 49—76.